\documentclass[aps,prd,preprint,superscriptaddress,showpacs,floatfix,nobibnotes]{revtex4}
\usepackage{mathrsfs}

\usepackage{bm}
\usepackage{graphicx}
\usepackage{longtable}
\usepackage{amsmath}

\begin{document}

\title{Radiative decay of hadronic molecule state for quarks }

\author{Xiaozhao Chen}\email{chen_xzhao@sina.com}
\email[corresponding author]{} \affiliation{Department of Foundational courses, Shandong University of Science and Technology, Taian,
271019, China}

\author{Xiaofu L\"{u}} \affiliation{Department of Physics,
Sichuan University, Chengdu, 610064, China} \affiliation{Institute
of Theoretical Physics, The Chinese Academy of Sciences, Beijing
100080, China} \affiliation{CCAST (World Laboratory), P.O. Box 8730,
Beijing  100080, China}

\author{Renbin Shi}
\affiliation{Department of Foundational courses, Shandong University of Science and Technology, Taian,
271019, China}

\author{Xiurong Guo}
 \affiliation{Department of Foundational courses, Shandong University of Science and Technology, Taian,
271019, China}

\author{Qingbiao Wang}
 \affiliation{Department of Foundational courses, Shandong University of Science and Technology, Taian,
271019, China}

\date{\today}

\begin{abstract}
Using the general form of the generalized Bethe-Salpeter wave functions for four-quark states describing the meson-meson molecular structure given in our previous work,   we obtain the general formulas for the decay widths of molecular states composed of two vector mesons with arbitrary spin and parity into two photons. Then this general formalism is applied to investigate the radiative two-photon decay of the obeserved \emph{X}(3915) state, where this exotic state \emph{X}(3915) is considered as a molecular state consisting of two heavy vector mesons $D^{*0}\bar{D}^{*0}$. The numerical result of decay mode $\emph{X}(3915)\rightarrow \gamma\gamma$ is consistent with the experimental values.
 \end{abstract}



\maketitle

\newpage

\parindent=20pt

\section{Introduction}
In the framework of quantum chromodynamics (QCD), beyond quark-antiquark ($q\bar q$) state there should be the other internal structures, such as tetraquark state and molecular state, etc, which have been used to interpret the exotic mesons \cite{ms:Swanson,ms:Torn,ms:Branz,ds:Maian1,ds:Maian2,ts:Ebert}. It is the most reasonable and fascinating to study the structure of exotic meson from QCD. In our previous works  \cite{mypaper4,mypaper5,mypaper6}, we have carefully investigated the molecular state composed of two vector mesons as far as possible from QCD and obtained the  general form of  generalized Bethe-Salpeter (GBS) wave functions of molecular states as four-quark states. This GBS wave function for four-quark state  was applied to evaluate the strong decay width of molecular state composed of two heavy vector mesons into a heavy meson plus a light meson in Ref \cite{mypaper6}, while the radiative decay of molecular state has still not been investigated. In this paper, we will emphatically investigate the two-photon decay of the molecular state composed of two vector mesons.

Different from the previous works \cite{ms:Swanson,ms:Torn,ms:Branz} about the hadronic molecule states, in our approach the vector mesons in molecular state are considered as the bound states composed of a quark and an antiquark. Because of the spontaneous breaking of chiral symmetry, the effective interaction Lagrangian at low energy QCD can be regarded as  the Lagrangian for the interaction of light mesons  with quarks. According to the effective theory at low energy QCD, we can investigate the exchanged meson interaction with the quarks in the vector meson and obtain the interaction kernel between two vector mesons. Solving the Bethe-Salpeter (BS) equation, we obtained the  masses and BS wave functions of molecular states composed of two vector mesons \cite{mypaper4,mypaper5}. From the molecule state model, we gave the GBS wave function for four-quark state \cite{mypaper6}. The GBS wave function derived from QCD is an essential prerequisite to accurately calculate the decay width of molecular state containing the strong and radiative decays.

When investigating the radiative decay of molecular state composed of two vector mesons, we still consider the internal structure of these vector mesons and the decay interaction is derived from the photon interaction with the quarks in these vector mesons. The photon-quark interaction can be described by the exact interaction Lagrangian $\mathscr{L}_I=i\frac{n}{3}e\bar q\gamma_\mu qA_\mu$, where the value of $n$ is determined by the flavor of quark field. In this work, we investigate the  radiative two-photon decay of the molecular state and  the lowest order approximation of this decay mode is the second order S-matrix element. Finally, considering this radiative decay interaction and using the GBS wave function, we can obtain the general formulas for the matrix elements of charge current between four-quark state and vacuum and evaluate the two-photon decay width of molecular state.

Then this approach is used to investigate a significant process: the radiative two-photon decay of \emph{X}(3915) \cite{X39151,Y39404}. The experimental data of \emph{X}(3915), once named \emph{Y}(3940), introduces a new challenge to the ordinary $c\bar c$ charmonium interpretation \cite{Y39403,mutlistate}. We assume that this exotic meson \emph{X}(3915) is a molecular state composed of two heavy vector mesons $D^{*0}\bar{D}^{*0}$ and evaluate the matrix elements between four-quark state and vacuum and the radiative decay width without an extra parameter.  From QCD, we comprehensively and systematically analyze the radiative two-photon decay of the molecular state composed of two heavy vector mesons and the calculated decay width $\Gamma(\emph{X}(3915)\rightarrow \gamma\gamma)$ is consistent with the experimental values, which provides a further verification for the molecular hypothesis of \emph{X}(3915). The Bethe-Salpeter theory is a relativistic theory for the two-body bound state in quantum field theory and our approach is an generalization of the BS theory, so our approach is in fact a nonperturbative method which can be applied to investigate arbitrary meson-meson molecular structure.

The structure of this article is as follows. In Sec. \ref{sec:BSWFFBS}  the GBS wave function of molecular state as a four-quark state is given. Section \ref{sec:BSME} gives the general formulas for the matrix elements of charge current between four-quark state and vacuum. In Sec. \ref{sec:DMX} our approach is used to investigate the decay mode $\emph{X}(3915)\rightarrow \gamma\gamma$ and this decay width is calculated. Our numerical result is presented in Sec. \ref{sec:nr} and we make some concluding remarks in Sec. \ref{sec:concl}.

\section{GBS WAVE FUNCTION FOR FOUR-QUARK STATE}\label{sec:BSWFFBS}
\begin{figure}[!htb] \centering
\includegraphics[scale=1,width=7cm]{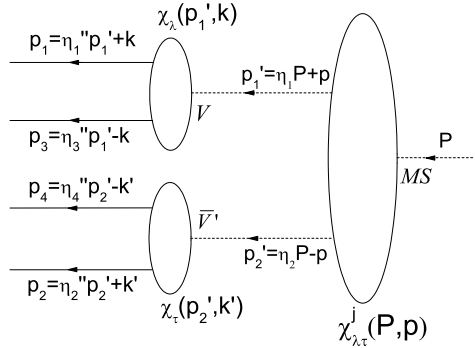}
\caption{\label{Fig1}  Generalized Bethe-Salpeter wave function for four-quark state in the momentum representation. The solid lines denote quark propagators; and the unfilled ellipses represent Bethe-Salpeter amplitudes.}
\end{figure}

 If a bound state with spin $j$ and parity $\eta_{P}$ is composed of four quarks, its GBS wave function can be defined as \cite{mypaper6}
\begin{equation}
\chi^j_P(x_1,x_3,x_4,x_2)=\langle 0|T\mathcal{Q}^C(x_1)\bar{\mathcal{Q}}^A(x_3)\mathcal{Q}^B(x_4)\bar{\mathcal{Q}}^D(x_2)|P\rangle,
\end{equation}
where $P$ is the momentum of the four-quark bound state, $\mathcal{Q}$ is the quark operator and its superscript is a flavor label. From translational invariance, this GBS wave function can be written as
\begin{equation}
\chi^j_{P}(x_1,x_3,x_4,x_2)=\frac{1}{(2\pi)^{3/2}}\frac{1}{\sqrt{2E(P)}}e^{iP\cdot X}\chi^j_{P}(X',x,x'),
\end{equation}
where $E(p)=\sqrt{\textbf{p}^2+m^2}$, $X=\eta_1(\eta_1''x_1+\eta_3''x_3)+\eta_2(\eta_4''x_4+\eta_2''x_2)$, $X'=(\eta_1''x_1+\eta_3''x_3)-(\eta_4''x_4+\eta_2''x_2)$, $x=x_1-x_3$, $x'=x_2-x_4$, $\eta_1+\eta_2=1$, $\eta''_{1,3}=m_{C,A}/(m_C+m_A)$, $\eta''_{2,4}=m_{D,B}/(m_D+m_B)$ and $m_{A,B,C,D}$ are the quark masses. Then making the Fourier transformation, we obtain the GBS wave function of four-quark bound state in the momentum representation
\begin{equation}\label{fourquarkBSWF}
\begin{split}
\chi^j_{P}(p_1,p_3,p_4,p_2)&=\int d^4x_1d^4x_3d^4x_4d^4x_2\chi^j_{P}(x_1,x_3,x_4,x_2)e^{-ip_1\cdot x_1}e^{-ip_3\cdot x_3}e^{-ip_4\cdot x_4}e^{-ip_2\cdot x_2}\\
&=\frac{1}{(2\pi)^{3/2}}\frac{1}{\sqrt{2E(P)}}(2\pi)^4\delta^{(4)}(P-p_1-p_3-p_4-p_2)\chi^j(P,p,k,k'),
\end{split}
\end{equation}
where $p_1, p_3, p_4, p_2$ are the momenta carried by the fields $\mathcal{Q}^C$, $\bar{\mathcal{Q}}^A$, $\mathcal{Q}^B$, $\bar{\mathcal{Q}}^D$; $p$, $k$, $k'$ are the conjugate variables to $X'$, $x$, $x'$, respectively; and $p=\eta_2(p_1+p_3)-\eta_1(p_4+p_2)$, $k=\eta_3''p_1-\eta_1''p_3$, $k'=\eta_4''p_2-\eta_2''p_4$. In the hadronic molecule structure, $p$ is the relative momentum between two mesons in molecular state, $k$ and $k'$ are the relative momenta between quark and antiquark in these two mesons, respectively, shown as in Fig. \ref{Fig1}. This work is aimed to investigate the molecular state composed of two vector mesons. In Fig. \ref{Fig1}, $V$ represents the vector meson with mass $M_1$ and $\bar{V'}$ represents the anti-particle of  vector meson $V'$ with mass $M_2$, $MS$ represents the vector-vector molecular state.

In Fig. \ref{Fig1}, there are three two-body systems in molecular state: a meson-meson bound state and two quark-antiquark bound states. We define the BS wave functions of these two-body systems as $\chi^j_P(p_1',p_2')$, $\chi_{p_1'}(p_1,p_3)$, $\chi_{p_2'}(p_4,p_2)$, respectively. According to the BS theory, the BS wave function for the bound state of two vector mesons has the form
\begin{equation}
\begin{split}
\chi^j_P(p_1',p_2')_{\lambda\tau}=\frac{1}{(2\pi)^{3/2}}\frac{1}{\sqrt{2E(P)}}(2\pi)^4\delta^{(4)}(P-p_1'-p_2')\chi^j_{\lambda\tau}(P,p),
\end{split}
\end{equation}
and the BS wave functions of two vector mesons are
\begin{equation}
\begin{split}
\chi_{p_1'}(p_1,p_3)_\lambda=\frac{1}{(2\pi)^{3/2}}\frac{1}{\sqrt{2E(p'_1)}}(2\pi)^4\delta^{(4)}(p_1'-p_1-p_3)\chi_\lambda(p_1',k),
\end{split}
\end{equation}
\begin{equation}
\begin{split}
\chi_{p_2'}(p_4,p_2)_\tau=\frac{1}{(2\pi)^{3/2}}\frac{1}{\sqrt{2E(p'_2)}}(2\pi)^4\delta^{(4)}(p_2'-p_4-p_2)\chi_\tau(p_2',k'),
\end{split}
\end{equation}
where $p_1'$ and $p_2'$ are the momenta of two vector mesons, respectively, $p_1'=\eta_1P+p$, $p_2'=\eta_2P-p$ and $\eta_{1,2}=M_{1,2}/(M_1+M_2)$. Applying the Feynman rules and comparing with Eq. (\ref{fourquarkBSWF}), we obtain the GBS wave function for four-quark state describing the molecular state composed of two vector mesons with arbitrary spin and definite parity \cite{mypaper6}
\begin{equation}\label{fourquarkBSWF1}
\begin{split}
\chi^j(P,p,k,k')=(2\pi)^8\chi_{\lambda}(p_1',k)\chi^j_{\lambda\tau}(P,p)\chi_\tau(p_2',k').
\end{split}
\end{equation}

In Ref. \cite{mypaper4}, we have given the general form of the BS wave functions for the bound states composed of two massive vector fields with spin $j$ and parity $\eta_{P}$, for $\eta_{P}=(-1)^j$,
\begin{equation}\label{jp0}
\chi_{\lambda\tau}^{j=0}(P,p)=\frac{1}{\mathcal{N}^j}(T^1_{\lambda\tau}\phi_1+T^2_{\lambda\tau}\phi_2),
\end{equation}
\begin{equation}\label{jp}
\chi_{\lambda\tau}^{j\neq0}(P,p)=\frac{1}{\mathcal{N}^j}\eta_{\mu_1\cdots\mu_j}[p_{\mu_1}\cdots
p_{\mu_j}(T^1_{\lambda\tau}\phi_1+T^2_{\lambda\tau}\phi_2)+T^3_{\lambda\tau}\phi_3+T^4_{\lambda\tau}\phi_4],
\end{equation}
and, for $\eta_{P}=(-1)^{j+1}$,
\begin{equation}\label{jm0}
\chi_{\lambda\tau}^{j=0}(P,p)=\frac{1}{\mathcal{N}^j}\epsilon_{\lambda\tau\xi\zeta}p_\xi
P_\zeta\psi_1,
\end{equation}
\begin{equation}\label{jm}
\chi_{\lambda\tau}^{j\neq0}(P,p)=\frac{1}{\mathcal{N}^j}\eta_{\mu_1\cdots\mu_j}(p_{\mu_1}\cdots
p_{\mu_j}\epsilon_{\lambda\tau\xi\zeta}p_\xi P_\zeta\psi_1+T^5_{\lambda\tau}\psi_2+T^6_{\lambda\tau}\psi_3+T^7_{\lambda\tau}\psi_4+T^8_{\lambda\tau}\psi_5),
\end{equation}
where $\mathcal{N}^j$ is the normalization, $\eta_{\mu_1\cdots\mu_j}$ is the polarization tensor describing the spin of the bound state, the subscripts $\lambda$ and $\tau$ are derived from these two vector fields, the independent tensor structures $T^i_{\lambda\tau}$ are given in Appendix A, $\phi_i(P\cdot p,p^2)$ and $\psi_i(P\cdot p,p^2)$ are independent scalar functions. For the vector mesons, the authors of Ref. \cite{BSE:Roberts1,BSE:Roberts3,BSE:Roberts4,BSE:Roberts5} have obtained their BS amplitudes in Euclidean space \cite{BSE:Roberts4,BSE:Roberts5}:
\begin{equation}
\begin{split}
\Gamma_{\lambda}^{V}(p'_1;k)=\frac{1}{\mathcal{N}^{V}}\bigg(\gamma_{\lambda}+p'_{1\lambda}\frac{\gamma\cdot
p'_1}{M_{V}^{2}}\bigg)\varphi_{V}(k^{2}),\\
\Gamma_{\tau}^{\bar V'}(p'_2;k')=\frac{1}{\mathcal{N}^{\bar V'}}\bigg(\gamma_{\tau}+p'_{2\tau}\frac{\gamma\cdot
p'_2}{M_{\bar V'}^{2}}\bigg)\varphi_{\bar V'}(k'^{2}),
\end{split}
\end{equation}
where $\Gamma_{\lambda}^{V}(p'_1;k)$ and $\Gamma_{\tau}^{\bar V'}(p'_2;k')$  are transverse ($p'_{1\lambda}\Gamma_{\lambda}^{V}(p'_1;k)=p'_{2\tau}\Gamma_{\tau}^{\bar V'}(p'_2;k')=0$), $\mathcal{N}^V$ and $\mathcal{N}^{\bar V'}$ are the normalizations, $\varphi_{V}(k^{2})$ and $\varphi_{\bar V}(k^{2})$ are scalar functions fixed by providing fits to the observables.  The BS wave functions of vector mesons are \cite{mypaper3}
\begin{equation}\label{BSwfvm}
\begin{split}
&\chi_\lambda(p_1',k)=\frac{-i}{\gamma^C\cdot p_1-im_C}\frac{1}{\mathcal{N}^{V}}\bigg(\gamma_\lambda+p_{1\lambda}'\frac{\gamma\cdot p_1'}{M_{V}^2}\bigg)\varphi_{V}(k^2)\frac{-i}{\gamma^A\cdot p_3-im_A},\\
&\chi_\tau(p_2',k')=\frac{-i}{\gamma^B\cdot p_4-im_B}\frac{1}{\mathcal{N}^{\bar V'}}\bigg(\gamma_\tau+p_{2\tau}'\frac{\gamma\cdot p_2'}{M_{\bar V'}^2}\bigg)\varphi_{\bar V'}(k'^2)\frac{-i}{\gamma^D\cdot p_2-im_D}.
\end{split}
\end{equation}

\section{GENERAL MATRIX ELEMENT BETWEEN FOUR-QUARK STATE AND VACUUM}\label{sec:BSME}
\begin{figure}[!htb] \centering
\includegraphics[scale=1,width=8cm]{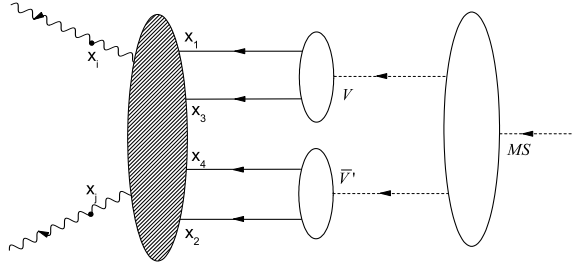}
\caption{\label{Fig2} General matrix element of charge current between four-quark state and vacuum in the coordinate representation. The filled ellipse represents the irreducible part of Green's function.}
\end{figure}

Because of the internal structure of the vector mesons in molecular state, we investigate the photon interaction with the quarks in vector mesons and consider that the quarks in a vector meson have different flavors in this work. The interaction Lagrangian for the coupling of the quarks to photon is
\begin{equation}
\begin{split}
&\mathscr{L}_I(x_i)=i\frac{n}{3}e\bar{\mathcal{Q}}(x_i) \gamma_\mu \mathcal{Q}(x_i)A_\mu(x_i)+i\frac{n'}{3}e\bar{\mathcal{Q}}'(x_i)\gamma_\mu \mathcal{Q}'(x_i)A_\mu(x_i),
\end{split}
\end{equation}
where $e$ is the electron charge, $\frac{e^2}{4\pi}=\frac{1}{137}$, the factors $n$ and $n'$ are determined by the flavor of quark, $\mathcal{Q}$ and $\mathcal{Q}'$ represent the quark field operators with different flavors. In Fig. \ref{Fig2} the  radiative two-photon decay of the molecular state is shown. The second order S-matrix element between molecular state and two photons can be obtained
\begin{equation}\label{Sme}
\begin{split}
&\langle \gamma\gamma|S^{(2)}|MS\rangle=\frac{(-i)^2}{2!}\int d^4x_i\int d^4x_j\langle \gamma\gamma|T\mathscr{H}_I(x_i)\mathscr{H}_I(x_j)|MS\rangle,\\
\end{split}
\end{equation}
and the lowest order transition matrix element for the radiative two-photon decay of the molecular state composed of two vector mesons is
\begin{equation}\label{Sme1}
\begin{split}
\langle \gamma\gamma|S^{(2)}|MS\rangle=&nn'\bigg(\frac{e}{3}\bigg)^2\int d^4x_i\int d^4x_j\\
&\times\langle \gamma\gamma|T\bar{\mathcal{Q}}(x_i)\gamma_\mu \mathcal{Q}(x_i)A_\mu(x_i)\bar{\mathcal{Q}}'(x_j)\gamma_\nu \mathcal{Q}'(x_j)A_\nu(x_j)|MS\rangle,\\
\end{split}
\end{equation}
where the factor $\frac{1}{2!}$ in Eq. (\ref{Sme}) has been cancelled after integration over $x_i$ and $x_j$. According to the S-matrix theory, we obtain
\begin{equation}\label{Sme2}
\begin{split}
\langle \gamma\gamma|S^{(2)}|MS\rangle=&nn'\bigg(\frac{e}{3}\bigg)^2\int d^4x_id^4x_j\\
&\times\frac{1}{(2\pi)^{3/2}}\frac{1}{\sqrt{2|\textbf{Q}'|}}\varepsilon_\mu^{\kappa'*}(Q')e^{-iQ'\cdot x_i}\frac{1}{(2\pi)^{3/2}}\frac{1}{\sqrt{2|\textbf{Q}|}}\varepsilon_\nu^{\kappa*}(Q)e^{-iQ\cdot x_j}\\
&\times\langle 0|T\bar{\mathcal{Q}}(x_i)\gamma_\mu \mathcal{Q}(x_i)\bar{\mathcal{Q}}'(x_j)\gamma_\nu \mathcal{Q}'(x_j)|P \rangle,
\end{split}
\end{equation}
where $Q$ and $Q'$ are the momenta of two photons in the final state, $Q=(\textbf{Q},i|\textbf{Q}|)$, $Q=(\textbf{Q}',i|\textbf{Q}'|)$, $\varepsilon_\nu^{\kappa=1,2}(Q)$ and $\varepsilon_\mu^{\kappa'=1,2}(Q')$ are their polarization vectors, respectively.

\begin{figure}[!htb] \centering
\includegraphics[scale=1,width=7cm]{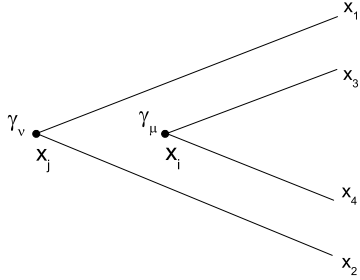}
\caption{\label{Fig3} The lowest order term of the two-particle irreducible Green's function.}
\end{figure}

Then the charge current in Eq. (\ref{Sme2}) is $J_\mu(x_i)J_\nu(x_j)=\bar{\mathcal{Q}}(x_i)\gamma_\mu \mathcal{Q}(x_i)\bar{\mathcal{Q}}'(x_j)\gamma_\nu \mathcal{Q}'(x_j)$. Applying Mandelstam's approach in quantum field theory \cite{Mandelstam}, we obtain the general formulas for the matrix elements of charge current $J_\mu J_\nu$ between four-quark state and vacuum
\begin{equation}\label{mefv}
\begin{split}
\langle 0|TJ_\mu(x_i)J_\nu(x_j)|P \rangle=&\int d^4x_1d^4x_3d^4x_4d^4x_2T(x_i,x_j;x_1,x_3,x_4,x_2)\chi^j_P(x_1,x_3,x_4,x_2),
\end{split}
\end{equation}
where $T(x_i,x_j;x_1,x_3,x_4,x_2)$ is the irreducible part of Green's function, shown as in Fig. \ref{Fig2}. It is straightforward to derive the two-particle irreducible Green's function
\begin{equation}
\begin{split}
T(x_i,x_j;x_1,x_3,x_4,x_2)=&\langle 0|T\bar{\mathcal{Q}}(x_i)\gamma_\mu \mathcal{Q}(x_i)\bar{\mathcal{Q}}'(x_j)\gamma_\nu \mathcal{Q}'(x_j)\\
&\times\bar{\mathcal{Q}}^C(x_1)\mathcal{Q}^A(x_3)\bar{\mathcal{Q}}^B(x_4)\mathcal{Q}^D(x_2)|0\rangle_T.
\end{split}
\end{equation}
Since the electromagnetic interaction does not change the quark flavor, we obtain
\begin{equation}\label{irGf}
\begin{split}
T(x_i,x_j;x_1,x_3,x_4,x_2)=&\langle 0|T\bar{\mathcal{Q}^A}(x_i)\gamma^A_\mu \mathcal{Q}^A(x_i)\bar{\mathcal{Q}}^C(x_j)\gamma^C_\nu \mathcal{Q}^C(x_j)\\
&\times\bar{\mathcal{Q}}^C(x_1)\mathcal{Q}^A(x_3)\bar{\mathcal{Q}}^A(x_4)\mathcal{Q}^C(x_2)|0\rangle_T,
\end{split}
\end{equation}
which can be calculated by means of perturbation theory. The lowest order term in the expansion of the right-hand side of Eq. (\ref{irGf}) is shown as Fig. \ref{Fig3} and the lowest order value of $T(x_i,x_j;x_1,x_3,x_4,x_2)$ is
\begin{equation}\label{irrec}
\begin{split}
T_0(x_i,x_j;x_1,x_3,x_4,x_2)=\delta^{(4)}(x_3-x_i)\gamma^A_\mu\delta^{(4)}(x_i-x_4)\delta^{(4)}(x_2-x_j)\gamma^C_\nu\delta^{(4)}(x_j-x_1).
\end{split}
\end{equation}
Substituting Eq. (\ref{irrec}) into (\ref{mefv}), we obtain the lowest order matrix elements of $J_\mu J_\nu$ between four-quark state and vacuum
\begin{equation}\label{mefv1}
\begin{split}
\langle 0|TJ_\mu(x_i)J_\nu(x_j)|P \rangle=&\int d^4x_1d^4x_3d^4x_4d^4x_2\delta^{(4)}(x_3-x_i)\gamma^A_\mu\delta^{(4)}(x_i-x_4)\\
&\times\delta^{(4)}(x_2-x_j)\gamma^C_\nu\delta^{(4)}(x_j-x_1)\chi^j_P(x_1,x_3,x_4,x_2)\\
=&\gamma^A_\mu\gamma^C_\nu\chi^j_P(x_j,x_i,x_i,x_j).
\end{split}
\end{equation}
From Eq. (\ref{fourquarkBSWF}), the result (\ref{mefv1}) can be transformed to
\begin{equation}\label{bsmep}
\begin{split}
\langle 0|TJ_\mu(x_i)J_\nu(x_j)|P \rangle=&\frac{1}{(2\pi)^{16}}\int d^4p_1d^4p_3d^4p_4d^4p_2\gamma^A_\mu\gamma^C_\nu\\
&\times\chi^j_P(p_1,p_3,p_4,p_2)e^{ip_1\cdot x_j}e^{ip_3\cdot x_i}e^{ip_4\cdot x_i}e^{ip_2\cdot x_j}\\
=&\frac{1}{(2\pi)^{12}}\frac{1}{(2\pi)^{3/2}}\frac{1}{\sqrt{2E(P)}}\int d^4p_1d^4p_3d^4p_4d^4p_2\gamma^A_\mu\gamma^C_\nu\\
&\times\delta^{(4)}(P-p_1-p_3-p_4-p_2)\chi^j(P,p,k,k')e^{i(p_1+p_2)\cdot x_j}e^{i(p_3+p_4)\cdot x_i},
\end{split}
\end{equation}
where $\chi^j(P,p,k,k')$ is the  GBS wave function for four-quark state expressed as Eq. (\ref{fourquarkBSWF1}), $p_1=\eta''_1(\eta_1P+p)+k$, $p_2=(\eta_2P-p)-[P-Q-\eta_3''(\eta_1 P+p)+k]$, $p_3=\eta''_3(\eta_1P+p)-k$, $p_4=P-Q-\eta_3''(\eta_1 P+p)+k$ and $k'=\eta_4''(\eta_2P-p)-[P-Q-\eta_3''(\eta_1 P+p)+k]$.

\begin{figure}[!htb] \centering
\includegraphics[scale=1,width=9cm]{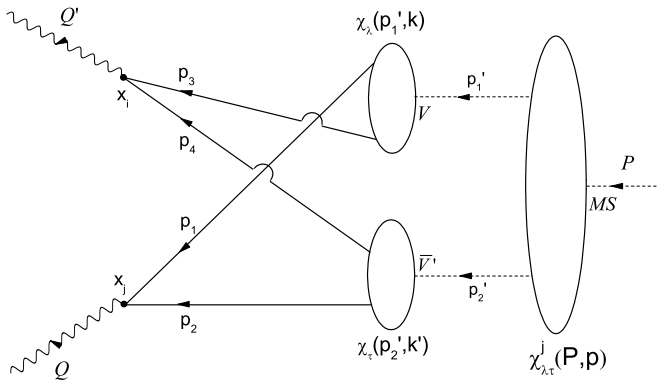}
\caption{\label{Fig4}  The lowest order matrix element between four-quark state and vacuum in the momentum representation.}
\end{figure}

Using  the matrix elements of $J_\mu J_\nu$ between four-quark state and vacuum in Eq. (\ref{bsmep}), we can obtain the computable form of the S-matrix element between molecular state and two photons in the momentum representation
\begin{equation}\label{Sme3}
\begin{split}
\langle \gamma\gamma|S^{(2)}|MS \rangle=&nn'\bigg(\frac{e}{3}\bigg)^2\frac{1}{(2\pi)^{3/2}}\frac{1}{\sqrt{2|\textbf{Q}'|}}\frac{1}{\sqrt{2|\textbf{Q}|}}\frac{1}{\sqrt{2E(P)}}\frac{1}{(2\pi)^{3}}\varepsilon_\mu^{\kappa'*}(Q')\varepsilon_\nu^{\kappa*}(Q)\\
&\times(2\pi)^{4}\delta^{(4)}(P-Q-Q')\int d^4kd^4p\frac{1}{(2\pi)^{8}}\gamma^A_\mu\gamma^C_\nu\chi^j(P,p,k,k').
\end{split}
\end{equation}
This matrix element is represented graphically in Fig. \ref{Fig4}. It is necessary to note that the two-photon decay of a molecular state has two possibilities by the interchange of two final photons and we find that the matrix element in Eq. (\ref{Sme3}) is invariant under the substitutions $Q\rightleftharpoons Q'$ and $\varepsilon^{\kappa}(Q)\rightleftharpoons\varepsilon^{\kappa'}(Q')$. Therefore, the total matrix element for the radiative two-photon decay of the molecular state composed of two vector mesons is
\begin{equation}
\begin{split}
\langle \gamma\gamma|S_{tot}^{(2)}|MS \rangle=2\langle \gamma\gamma|S^{(2)}|MS \rangle.
\end{split}
\end{equation}

\section{THE DECAY MODE $\bm{\emph{X}(3915)\rightarrow \gamma\gamma}$}\label{sec:DMX}
As an application, we investigate the radiative two-photon decay of $\emph{X}(3915)$ in this section. Here, the exotic state $\emph{X}(3915)$, once named \emph{Y}(3940), is considered as a S-wave molecule state consisting of two heavy vector mesons $D^{*0}$ and $\bar{D}^{*0}$. In Ref. \cite{mypaper4} we have obtained the mass and BS wave function of this molecular state and deduced that the spin-parity quantum numbers of the $\emph{X}(3915)$ system are $0^+$. In Fig. \ref{Fig4}, $V$ and $\bar V'$ become $D^{*0}$ and $\bar{D}^{*0}$, respectively. The heavy vector meson $D^{*0}$ in molecular state is composed of a heavy quark $c$ and a light antiquark $\bar u$, and the factors in coupling constants become $n=n'=2$. In Eqs. (\ref{BSwfvm}) and (\ref{Sme3}), the flavor labels $A=B$ and $C=D$ represent u-quark and c-quark, respectively.  Then the matrix element for this decay process becomes
\begin{equation}\label{bsmep2}
\begin{split}
\langle \gamma\gamma|S_{tot}^{(2)}|X(3915) \rangle=&2\bigg(\frac{2e}{3}\bigg)^2\frac{1}{(2\pi)^{3/2}}\frac{1}{\sqrt{2|\textbf{Q}'|}}\frac{1}{\sqrt{2|\textbf{Q}|}}\frac{1}{\sqrt{2E(P)}}\frac{1}{(2\pi)^{3}}\varepsilon_\mu^{\kappa'*}(Q')\varepsilon_\nu^{\kappa*}(Q)\\
&\times(2\pi)^{4}\delta^{(4)}(P-Q-Q')\int d^4kd^4p\frac{1}{(2\pi)^{8}}\gamma^A_\mu\gamma^C_\nu\chi^j(P,p,k,k').
\end{split}
\end{equation}
The GBS wave function for four-quark state is
\begin{equation}\label{fourquarkBSWF2}
\begin{split}
&\chi^{j=0}(P,p,k,k')\\
&=(2\pi)^8\chi_\lambda(p_1',k)\chi^{0^+}_{\lambda\tau}(P,p)\chi_\tau(p_2',k')\\
&=(2\pi)^8\frac{-i}{\gamma^C\cdot p_1-im_c}\frac{1}{\mathcal{N}^{D^{*0}}}\bigg(\gamma_\lambda+p_{1\lambda}'\frac{\gamma\cdot p_1'}{M_{D^{*0}}^2}\bigg)\varphi_{D^{*0}}(k^2)\frac{-i}{\gamma^A\cdot p_3-im_u}\frac{1}{\mathcal{N}^{0^+}}(T^1_{\lambda\tau}\mathcal{F}_1+T^2_{\lambda\tau}\mathcal{F}_2)\\
&~~~~\times\frac{-i}{\gamma^A\cdot p_4-im_u}\frac{1}{\mathcal{N}^{\bar D^{*0}}}\bigg(\gamma_\tau+p_{2\tau}'\frac{\gamma\cdot p_2'}{M_{\bar D^{*0}}^2}\bigg)\varphi_{\bar D^{*0}}(k'^2)\frac{-i}{\gamma^C\cdot p_2-im_c},
\end{split}
\end{equation}
where $m_{c,u}$ are the constituent quark masses, $\varphi_{D^{*0}}(k^{2})=\varphi_{\bar D^{*0}}(k^{2})=exp(-k^{2}/\omega_{D^{*0}}^{2})$, $\omega_{D^{*0}}$=1.50GeV \cite{BSE:Roberts5} and the momentum of this bound
state is set as $P=(0,0,0,iM)$ in the rest frame.  As in heavy-quark effective theory (HQET) \cite{hqet}, we consider that the heaviest quark carries all the heavy-meson momentum and these momenta in Eq. (\ref{bsmep}) become
\begin{equation}\label{momenta}
\begin{split}
&p_1=P/2+p+k,~~p_2=Q-P/2-p-k,~~p_3=-k,~~p_4=-k',\\
&p_1'=P/2+p,~~p_2'=P/2-p,~~k'=Q-P-k.
\end{split}
\end{equation}

Now, we determine these normalizations $\mathcal{N}^{0^+}$ and $\mathcal{N}^{D^{*0}}$.  In Ref. \cite{mypaper4} the BS equation for the molecular state composed of two heavy vector mesons is treated in the ladder approximation and the BS wave function of the molecular state $D^{*0}$$\bar{D}^{*0}$ with $0^+$ is obtained
\begin{equation}
\begin{split}
\chi_{\lambda\tau}^{0^+}(P,p)=\frac{1}{\mathcal{N}^{0^+}}[T^1_{\lambda\tau}\mathcal{F}_1(P\cdot p,p^2)+T^2_{\lambda\tau}\mathcal{F}_2(P\cdot p,p^2)].
\end{split}
\end{equation}
The reduced normalization condition for $\chi_{\lambda\tau}^{0^+}(P,p)$ is
\begin{equation}
\begin{split}
\frac{-i}{(2\pi)^4}\int d^4p\bar\chi_{\lambda\tau}(P,p)\frac{\partial}{\partial P_0}[\Delta_{F\lambda\lambda'}(P/2+p)^{-1}\Delta_{F\tau\tau'}(P/2-p)^{-1}]\chi_{\lambda'\tau'}(P,p)=2P_0,
\end{split}
\end{equation}
where $\Delta_{F\beta\alpha'}(p)^{-1}$ is the inverse propagator for the vector field with mass $m$, $\Delta_{F\beta\alpha'}(p)^{-1}=i(\delta_{\beta\alpha'}-\frac{p_{\beta}
p_{\alpha'}}{p^2+m^2})(p^2+m^2)$ \cite{mypaper6}. These scalar functions $\mathcal{F}_1$ and $\mathcal{F}_2$ should satisfy two individual equations \cite{mypaper4}
\begin{equation*}
\mathcal{F}_1(P\cdot p,p^2)
=\frac{1}{(P/2+p)^2+M_1^2-i\epsilon}\frac{1}{(P/2-p)^2+M_2^2-i\epsilon}
\int\frac{d^4q'}{(2\pi)^4}V_{1}(p,q';P)\mathcal{F}_1(P\cdot
q',q'^2),
\end{equation*}
\begin{equation*}
\begin{split}
(P/2-p)^2\mathcal{F}_2(P\cdot p,p^2)
=&\frac{1}{(P/2+p)^2+M_1^2-i\epsilon}\frac{1}{(P/2-p)^2+M_2^2-i\epsilon}\\
&\times\int\frac{d^4q'}{(2\pi)^4}V_{2}(p,q';P)(P/2-q')^2\mathcal{F}_2(P\cdot
q',q'^2),
\end{split}
\end{equation*}
where $V_{1}(p,q';P)$ and $V_{2}(p,q';P)$ are derived from the interaction kernel between $D^{*0}$$\bar{D}^{*0}$. Solving these two equations in instantaneous approximation, we obtained the wave functions $\Psi_1^{0^+}(\textbf{p})=\int dp_0\mathcal{F}_1(P\cdot p,p^2)$
and $\Psi_2^{0^+}(\textbf{p})=\int dp_0(P/2-p)^2\mathcal{F}_2(P\cdot p,p^2)$ in Ref. \cite{mypaper4}. To fix the normalization $\mathcal{N}^{0^+}$, we require the scalar functions $\mathcal{F}_1$ and $\mathcal{F}_2$, which can be obtained from $\Psi_1^{0^+}$ and $\Psi_2^{0^+}$, respectively. From Eq. (\ref{BSwfvm}), the BS wave function of $D^{*0}$ meson can be written as
\begin{equation}\label{D0BSwf}
\begin{split}
&\chi_\lambda(K,k)=\frac{-i}{\gamma\cdot(K+k)-im_c}\frac{1}{\mathcal{N}^{D^{*0}}}\bigg(\gamma_\lambda+K_{\lambda}\frac{\gamma\cdot K}{M_{D^{*0}}^2}\bigg)\varphi_{D^{*0}}(k^2)\frac{-i}{\gamma\cdot(-k)-im_u},
\end{split}
\end{equation}
where $K$ is set as the momentum of the heavy meson in the rest frame, $k$ denotes the relative momentum between c-quark and u-antiquark,  $k$ and $K$ are not the momenta presented in the decay process. The authors of Refs. \cite{BSE:Roberts4,BSE:Roberts5} also employed the ladder approximation to solve the BS equation for quark-antiquark state, and the reduced normalization condition for the BS wave function of $D^{*0}$ meson given by Eq. (\ref{D0BSwf}) is
\begin{equation}
\begin{split}
\frac{-i}{(2\pi)^4}\frac{1}{3}\int d^4k\bar\chi_{\lambda}(K,k)\frac{\partial}{\partial K_0}[S(K+k)^{-1}]S(-k)^{-1}\chi_\lambda(K,k)=2K_0,
\end{split}
\end{equation}
where the factor $1/3$ appears because of the sum of three transverse directions.

Then the lowest order transition matrix element for the radiative two-photon decay of $\emph{X}(3915)$  expressed as Eq. (\ref{bsmep2}) can be calculated
\begin{equation}
\begin{split}
\langle \gamma\gamma|S_{tot}^{(2)}|X(3915) \rangle
=&2\bigg(\frac{2e}{3}\bigg)^2\frac{1}{(2\pi)^{3/2}}\frac{1}{\sqrt{2|\textbf{Q}'|}}\frac{1}{\sqrt{2|\textbf{Q}|}}\frac{1}{\sqrt{2E(P)}}\\
&\times\frac{1}{(2\pi)^{3}}\varepsilon_\mu^{\kappa'*}(Q')\varepsilon_\nu^{\kappa*}(Q)(2\pi)^{4}\delta^{(4)}(P-Q-Q')\mathcal{M}_{\nu\mu},
\end{split}
\end{equation}
where
\begin{equation}\label{MEcurr}
\begin{split}
\mathcal{M}_{\nu\mu}&=\int d^4kd^4p\frac{1}{p_1^2+m^2_c}\frac{1}{N^{D^{*0}}}\frac{\varphi_{D^{*0}}(k^2)}{k^2+m^2_u}\frac{1}{N^{\bar D^{*0}}}\frac{\varphi_{\bar D^{*0}}(k'^2)}{k'^2+m^2_u}\frac{1}{p_2^2+m^2_c}\frac{1}{\mathcal{N}^{0^+}}\\
&~~~~\times Tr\bigg\{\gamma_\nu[\gamma\cdot p_1+im_c]\{[(p_1'\cdot p_2')\gamma_\tau-(\gamma\cdot p_2')p_{1\tau}']\phi_1(p)\\
&~~~~+[p_1'^2p_2'^2\gamma_{\tau}+(p_1'\cdot p_2')(\gamma\cdot p_1')p_{2\tau}'-p_2'^2(\gamma\cdot p_1')p_{1\tau}'-p_1'^2(\gamma\cdot p_2')p_{2\tau}']\phi_2(p)\}\\
&~~~~\times[\gamma\cdot(-k)+im_u]\gamma_\mu[\gamma\cdot(-k')+im_u]\gamma_\tau[\gamma\cdot p_2+im_c]\bigg\}.
 \end{split}
\end{equation}
In Eq. (\ref{MEcurr}) the trace of the product of 8 $\gamma$-matrices contains 105 terms and the resulting expression has been given in Appendix B of Ref. \cite{mypaper6}. In our approach, the $p$ integral is computed in instantaneous approximation. To calculate this tensor $\mathcal{M}_{\nu\mu}$, we have given a simple method in Ref. \cite{mypaper6}. It is obvious that the tensor $\mathcal{M}_{\nu\mu}$ only depends on $P$ and $Q$, so in Minkowski space $\mathcal{M}_{\nu\mu}$ can be expressed  as
\begin{equation}\label{MEcurrLT}
\begin{split}
&\mathcal{M}_{\nu\mu}=g_{\nu\mu}f_1(P,Q)+P_\nu Q_\mu f_2(P,Q)+P_\nu P_\mu f_3(P,Q)+Q_\nu P_\mu f_4(P,Q)+Q_\nu Q_\mu f_5(P,Q),
\end{split}
\end{equation}
where $f_i(P,Q)$ are the scalar functions. The above expression is multiplied by these tensor structures $g_{\nu\mu}$, $P_\nu Q_\mu$, $P_\nu P_\mu$, $Q_\nu P_\mu$, $Q_\nu Q_\mu$, respectively; and a set of equations is obtained
\begin{equation}
\begin{split}
g_{\nu\mu}\mathcal{M}_{\nu\mu}=h_1&=4f_1+(P\cdot Q)f_2+P^2f_3+(P\cdot Q)f_4+Q^2f_5,\\
P_\nu Q_\mu\mathcal{M}_{\nu\mu}=h_2&=(P\cdot Q)f_1+P^2Q^2f_2+P^2(P\cdot Q)f_3+(P\cdot Q)^2f_4+Q^2(P\cdot Q)f_5,\\
P_\nu P_\mu\mathcal{M}_{\nu\mu}=h_3&=P^2f_1+P^2(P\cdot Q)f_2+P^2P^2f_3+P^2(P\cdot Q)f_4+(P\cdot Q)^2f_5,\\
Q_\nu P_\mu\mathcal{M}_{\nu\mu}=h_4&=(P\cdot Q)f_1+(P\cdot Q)^2f_2+P^2(P\cdot Q)f_3+Q^2P^2f_4+Q^2(P\cdot Q) f_5,\\
Q_\nu Q_\mu\mathcal{M}_{\nu\mu}=h_5&=Q^2f_1+Q^2(P\cdot Q)f_2+(P\cdot Q)^2f_3+Q^2(P\cdot Q)f_4+Q^2Q^2f_5,
\end{split}
\end{equation}
where $h_i$ are numbers. Subsequently, we numerically calculate $h_i$ and solve this set of equations. The values of $f_i$ can be obtained.

Finally, we obtain the $\emph{X}(3915)\rightarrow \gamma\gamma$ decay width
\begin{equation}\label{dw}
\begin{split}
\Gamma=&\int d^3Qd^3Q'(2\pi)^4\delta^{(4)}(P-Q-Q')4\bigg(\frac{2e}{3}\bigg)^4\frac{1}{2|\textbf{Q}'|}\frac{1}{2|\textbf{Q}|}\frac{1}{2E(P)}\frac{1}{(2\pi)^{6}}\\
&\times\sum_{\kappa'=1}^2\sum_{\kappa=1}^2|\varepsilon_\mu^{\kappa'}(Q')\varepsilon_\nu^{\kappa}(Q)\mathcal{M}_{\nu\mu}|^2\\
=&\frac{1}{\pi}\bigg(\frac{2e}{3}\bigg)^4\frac{1}{M}f_1^*f_1,
 \end{split}
\end{equation}
where $P=(0,0,0,iM)$, $Q=(\textbf{Q}_\gamma,i|\textbf{Q}_\gamma|)$, $Q'=(-\textbf{Q}_\gamma,i|\textbf{Q}_\gamma|)$, and $|\textbf{Q}_\gamma|=M/2$.  To derive Eq. (\ref{dw}), we have used the transverse condition of electromagnetic field $\varepsilon^\kappa(Q)\cdot Q=0$ and the completeness relation.

\section{Numerical result}\label{sec:nr}
The constituent quark masses $m_{u}=0.33$GeV, $m_{c}=1.55$GeV, the meson mass $M_{D^{*0}}=2.007$GeV \cite{PDG2016}. Our numerical result for the radiative two-photon decay of the exotic state $\emph{X}(3915)$ is $\Gamma(\emph{X}(3915)\rightarrow \gamma\gamma)=36$keV. The strong decay width of the $\emph{X}$ state has been calculated in our previous work \cite{mypaper6}, and the result is $\Gamma(\emph{X}(3915)\rightarrow J/\psi\omega)=66$MeV. The value of the product of these two calculated decay widths, $\Gamma(\emph{X}\rightarrow \gamma\gamma)\times\Gamma(\emph{X}\rightarrow J/\psi\omega)$, is compatible with the experimental data. In experiments \cite{X39151,Y39404},  the product of the strong $\emph{X}(3915)\rightarrow J/\psi\omega$ and radiative $\emph{X}(3915)\rightarrow \gamma\gamma$ decay widths is of order $10^3$keV$^2$.

To investigate the radiative decay of molecular state, we consider the photon interaction with the quarks, which can be described by the exact interaction Lagrangian. The only parameter $\omega_{D^{*0}}$ in the BS amplitude of heavy vector meson is not an adjustable  parameter, which is determined by providing fits to observables.  By doing the numerical calculation, we obtain the decay width $\Gamma(\emph{X}(3915)\rightarrow \gamma\gamma)$ within the range of experimental value. Therefore, this work provides a further verification for the molecular hypothesis of \emph{X}(3915) and predicts the exact value of the radiative two-photon decay width $\Gamma(\emph{X}(3915)\rightarrow \gamma\gamma)$.

Up to now, a systematical and accurate theoretical approach from QCD to investigate the molecular state composed of two heavy vector mesons has been established. Applying the general form of the Bethe-Salpeter wave functions for the bound states composed of two vector fields, we calculated the mass of the molecular state $D^{*0}\bar D^{*0}$ with $0^+$ and obtained its BS wave function \cite{mypaper4}. Then using the general form of the GBS wave functions for four-quark states describing the meson-meson molecular structure, we calculated the strong $\emph{X}(3915)\rightarrow J/\psi\omega$ decay width \cite{mypaper6}. In this work, we investigate the radiative two-photon decay of the molecular state.

\section{Conclusion}\label{sec:concl}
The general form of the GBS wave functions for four-quark states describing the molecular structure is applied to investigate the radiative two-photon decay of the molecular state composed of two vector mesons and the general formulas for the two-photon decay widths of molecular states is obtained. Then assuming that the exotic state \emph{X}(3915) is a molecular state of $D^{*0}\bar D^{*0}$, we carefully consider the internal structure of the vector mesons in molecular state and numerically calculate the decay width $\Gamma(\emph{X}(3915)\rightarrow \gamma\gamma)$, which is compatible with experiments. From QCD, we have  systematically investigated the strong and radiative decay modes of the molecular state composed of two heavy vector mesons. In the future, to contain more inspiration of QCD, we will introduce the nonperturbative contribution from the vacuum condensates into the GBS wave function for four-quark state and the irreducible part of Green's function.

\begin{acknowledgements}
This work was supported by the National Natural Science Foundation of China under Grants No. 11705104 and 11801323, and Shandong Provincial Natural Science Foundation, China under Grants No. ZR2016AQ19 and ZR2016AM31, and a project of Shandong Province Higher Education Science and Technology program under Grants No. J17KB130 and J18KA227, and SDUST Research Fund under Grant No. 2018TDJH101.
\end{acknowledgements}

\appendix

\section{The tensor structures in the general form of the BS wave functions}\label{app1}
The tensor structures in Eqs. (\ref{jp0}), (\ref{jp}), (\ref{jm0}), (\ref{jm}) are given below \cite{mypaper4}
\begin{equation*}
T_{\lambda\tau}^1=(\eta_1\eta_2P^2-\eta_1P\cdot p+\eta_2P\cdot p-p^2)g_{\lambda\tau}-(\eta_1\eta_2P_{\lambda}P_{\tau}+\eta_2P_{\lambda}p_{\tau}-\eta_1p_{\lambda}P_{\tau}-p_{\lambda}p_{\tau}),
\end{equation*}
\begin{equation*}
\begin{split}
T_{\lambda\tau}^2=&(\eta_1^2P^2+2\eta_1P\cdot p+p^2)(\eta_2^2P^2-2\eta_2P\cdot p+p^2)g_{\lambda\tau}\\
&+(\eta_1\eta_2P^2-\eta_1P\cdot p+\eta_2P\cdot p-p^2)(\eta_1\eta_2P_{\lambda}P_{\tau}-\eta_1P_{\lambda}p_{\tau}+\eta_2p_{\lambda}P_{\tau}-p_{\lambda}p_{\tau})\\
&-(\eta_2^2P^2-2\eta_2P\cdot p+p^2)(\eta_1 ^2P_{\lambda}P_{\tau}+\eta_1P_{\lambda}p_{\tau}+\eta_1p_{\lambda}P_{\tau}+p_{\lambda}p_{\tau})\\
&-(\eta_1^2P^2+2\eta_1P\cdot p+p^2)(\eta_2 ^2P_{\lambda}P_{\tau}-\eta_2P_{\lambda}p_{\tau}-\eta_2p_{\lambda}P_{\tau}+p_{\lambda}p_{\tau}),
\end{split}
\end{equation*}
\begin{equation*}
\begin{split}
T_{\lambda\tau}^3=&\frac{1}{j!}p_{\{\mu_2}\cdots
p_{\mu_j}g_{\mu_1\}\lambda}(\eta_1^2P^2+2\eta_1P\cdot p+p^2)[(\eta_2^2P^2-2\eta_2P\cdot p+p^2)(\eta_1P+p)_{\tau}\\
&-(\eta_1\eta_2P^2-\eta_1P\cdot p+\eta_2P\cdot p-p^2)(\eta_2P-p)_{\tau}]\\
&-p_{\mu_1}\cdots
p_{\mu_j}[(\eta_2^2P^2-2\eta_2P\cdot p+p^2)(\eta_1 ^2P_{\lambda}P_{\tau}+\eta_1P_{\lambda}p_{\tau}+\eta_1p_{\lambda}P_{\tau}+p_{\lambda}p_{\tau})\\
&-(\eta_1\eta_2P^2-\eta_1P\cdot p+\eta_2P\cdot p-p^2)(\eta_1\eta_2P_{\lambda}P_{\tau}-\eta_1P_{\lambda}p_{\tau}+\eta_2p_{\lambda}P_{\tau}-p_{\lambda}p_{\tau})],
\end{split}
\end{equation*}
\begin{equation*}
\begin{split}
T_{\lambda\tau}^4=&\frac{1}{j!}p_{\{\mu_2}\cdots
p_{\mu_j}g_{\mu_1\}\tau}(\eta_2^2P^2-2\eta_2P\cdot p+p^2)[(\eta_1\eta_2P^2-\eta_1P\cdot p\\
&+\eta_2P\cdot p-p^2)(\eta_1P+p)_{\lambda}-(\eta_1^2P^2+2\eta_1P\cdot p+p^2)(\eta_2P-p)_{\lambda}]\\
&-p_{\mu_1}\cdots
p_{\mu_j}[(\eta_1^2P^2+2\eta_1P\cdot p+p^2)(\eta_2 ^2P_{\lambda}P_{\tau}-\eta_2P_{\lambda}p_{\tau}-\eta_2p_{\lambda}P_{\tau}+p_{\lambda}p_{\tau})\\
&-(\eta_1\eta_2P^2-\eta_1P\cdot p+\eta_2P\cdot p-p^2)(\eta_1\eta_2P_{\lambda}P_{\tau}-\eta_1P_{\lambda}p_{\tau}+\eta_2p_{\lambda}P_{\tau}-p_{\lambda}p_{\tau})],
\end{split}
\end{equation*}
\begin{equation*}
\begin{split}
T_{\lambda\tau}^5=&(\eta_2P\cdot p-\eta_1P\cdot p-2p^2)p_{\{\mu_2}\cdots
p_{\mu_j}\epsilon_{\mu_1\}\lambda\tau\xi}p_\xi\\
&+(2\eta_1\eta_2P\cdot
p+\eta_2p^2-\eta_1p^2)p_{\{\mu_2}\cdots
p_{\mu_j}\epsilon_{\mu_1\}\lambda\tau\xi}P_\xi\\
&+p_{\{\mu_2}\cdots
p_{\mu_j}\epsilon_{\mu_1\}\lambda\xi\zeta}p_\xi P_\zeta
p_\tau+p_{\{\mu_2}\cdots
p_{\mu_j}\epsilon_{\mu_1\}\tau\xi\zeta}p_\xi P_\zeta
p_\lambda,
\end{split}
\end{equation*}
\begin{equation*}\
\begin{split}
T_{\lambda\tau}^6=&(P\cdot p)p_{\{\mu_2}\cdots
p_{\mu_j}\epsilon_{\mu_1\}\lambda\tau\xi}p_\xi-p^2p_{\{\mu_2}\cdots
p_{\mu_j}\epsilon_{\mu_1\}\lambda\tau\xi}P_\xi\\
&+p_{\{\mu_2}\cdots
p_{\mu_j}\epsilon_{\mu_1\}\lambda\xi\zeta}p_\xi P_\zeta
p_\tau-p_{\{\mu_2}\cdots
p_{\mu_j}\epsilon_{\mu_1\}\tau\xi\zeta}p_\xi P_\zeta
p_\lambda,
\end{split}
\end{equation*}
\begin{equation*}
\begin{split}
T_{\lambda\tau}^7=&(\eta_2P^2-\eta_1P^2-2P\cdot p)p_{\{\mu_2}\cdots
p_{\mu_j}\epsilon_{\mu_1\}\lambda\tau\xi}p_\xi\\
&+(2\eta_1\eta_2P^2+\eta_2P\cdot p-\eta_1P\cdot p)p_{\{\mu_2}\cdots
p_{\mu_j}\epsilon_{\mu_1\}\lambda\tau\xi}P_\xi\\
&+p_{\{\mu_2}\cdots
p_{\mu_j}\epsilon_{\mu_1\}\lambda\xi\zeta}p_\xi P_\zeta
P_\tau+p_{\{\mu_2}\cdots
p_{\mu_j}\epsilon_{\mu_1\}\tau\xi\zeta}p_\xi P_\zeta
P_\lambda,
\end{split}
\end{equation*}
\begin{equation*}
\begin{split}
T_{\lambda\tau}^8=&P^2p_{\{\mu_2}\cdots
p_{\mu_j}\epsilon_{\mu_1\}\lambda\tau\xi}p_\xi-(P\cdot p)p_{\{\mu_2}\cdots
p_{\mu_j}\epsilon_{\mu_1\}\lambda\tau\xi}P_\xi\\
&+p_{\{\mu_2}\cdots
p_{\mu_j}\epsilon_{\mu_1\}\lambda\xi\zeta}p_\xi P_\zeta
P_\tau-p_{\{\mu_2}\cdots
p_{\mu_j}\epsilon_{\mu_1\}\tau\xi\zeta}p_\xi P_\zeta
P_\lambda.
\end{split}
\end{equation*}

\bibliographystyle{apsrev}
\bibliography{ref}

\begin{thebibliography}{21}
\expandafter\ifx\csname natexlab\endcsname\relax\def\natexlab#1{#1}\fi
\expandafter\ifx\csname bibnamefont\endcsname\relax
  \def\bibnamefont#1{#1}\fi
\expandafter\ifx\csname bibfnamefont\endcsname\relax
  \def\bibfnamefont#1{#1}\fi
\expandafter\ifx\csname citenamefont\endcsname\relax
  \def\citenamefont#1{#1}\fi
\expandafter\ifx\csname url\endcsname\relax
  \def\url#1{\texttt{#1}}\fi
\expandafter\ifx\csname urlprefix\endcsname\relax\def\urlprefix{URL }\fi
\providecommand{\bibinfo}[2]{#2}
\providecommand{\eprint}[2][]{\url{#2}}

\bibitem[{\citenamefont{Swanson}(2004)}]{ms:Swanson}
\bibinfo{author}{\bibfnamefont{E.~S.} \bibnamefont{Swanson}},
  \bibinfo{journal}{Phys. Lett. B} \textbf{\bibinfo{volume}{588}},
  \bibinfo{pages}{189} (\bibinfo{year}{2004}).

\bibitem[{\citenamefont{T{\"o}rnqvist}(2004)}]{ms:Torn}
\bibinfo{author}{\bibfnamefont{N.~A.} \bibnamefont{T{\"o}rnqvist}},
  \bibinfo{journal}{Phys. Lett. B} \textbf{\bibinfo{volume}{590}},
  \bibinfo{pages}{209} (\bibinfo{year}{2004}).

\bibitem[{\citenamefont{Branz et~al.}(2009)\citenamefont{Branz, Gutsche, and
  Lyubovitskij}}]{ms:Branz}
\bibinfo{author}{\bibfnamefont{T.}~\bibnamefont{Branz}},
  \bibinfo{author}{\bibfnamefont{T.}~\bibnamefont{Gutsche}}, \bibnamefont{and}
  \bibinfo{author}{\bibfnamefont{V.~E.} \bibnamefont{Lyubovitskij}},
  \bibinfo{journal}{Phys. Rev. D} \textbf{\bibinfo{volume}{80}},
  \bibinfo{pages}{054019} (\bibinfo{year}{2009}).

\bibitem[{\citenamefont{Maiani et~al.}(2005)\citenamefont{Maiani, Piccinini,
  Polosa, and Riquer}}]{ds:Maian1}
\bibinfo{author}{\bibfnamefont{L.}~\bibnamefont{Maiani}},
  \bibinfo{author}{\bibfnamefont{F.}~\bibnamefont{Piccinini}},
  \bibinfo{author}{\bibfnamefont{A.~D.} \bibnamefont{Polosa}},
  \bibnamefont{and} \bibinfo{author}{\bibfnamefont{V.}~\bibnamefont{Riquer}},
  \bibinfo{journal}{Phys. Rev. D} \textbf{\bibinfo{volume}{71}},
  \bibinfo{pages}{014028} (\bibinfo{year}{2005}).

\bibitem[{\citenamefont{Maiani et~al.}(2007)\citenamefont{Maiani, Polosa, and
  Riquer}}]{ds:Maian2}
\bibinfo{author}{\bibfnamefont{L.}~\bibnamefont{Maiani}},
  \bibinfo{author}{\bibfnamefont{A.~D.} \bibnamefont{Polosa}},
  \bibnamefont{and} \bibinfo{author}{\bibfnamefont{V.}~\bibnamefont{Riquer}},
  \bibinfo{journal}{Phys. Rev. Lett.} \textbf{\bibinfo{volume}{99}},
  \bibinfo{pages}{182003} (\bibinfo{year}{2007}).

\bibitem[{\citenamefont{Ebert et~al.}(2006)\citenamefont{Ebert, Faustov, and
  Galkin}}]{ts:Ebert}
\bibinfo{author}{\bibfnamefont{D.}~\bibnamefont{Ebert}},
  \bibinfo{author}{\bibfnamefont{R.~N.} \bibnamefont{Faustov}},
  \bibnamefont{and} \bibinfo{author}{\bibfnamefont{V.~O.}
  \bibnamefont{Galkin}}, \bibinfo{journal}{Phys. Lett. B}
  \textbf{\bibinfo{volume}{634}}, \bibinfo{pages}{214} (\bibinfo{year}{2006}).

\bibitem[{\citenamefont{Chen and L{\"u}}(2015)}]{mypaper4}
\bibinfo{author}{\bibfnamefont{X.}~\bibnamefont{Chen}} \bibnamefont{and}
  \bibinfo{author}{\bibfnamefont{X.}~\bibnamefont{L{\"u}}},
  \bibinfo{journal}{Eur. Phys. J. C} \textbf{\bibinfo{volume}{75}},
  \bibinfo{pages}{98} (\bibinfo{year}{2015}).

\bibitem[{\citenamefont{Chen et~al.}(2016)\citenamefont{Chen, L{\"u}, Shi, and
  Guo}}]{mypaper5}
\bibinfo{author}{\bibfnamefont{X.}~\bibnamefont{Chen}},
  \bibinfo{author}{\bibfnamefont{X.}~\bibnamefont{L{\"u}}},
  \bibinfo{author}{\bibfnamefont{R.}~\bibnamefont{Shi}}, \bibnamefont{and}
  \bibinfo{author}{\bibfnamefont{X.}~\bibnamefont{Guo}},
  \bibinfo{journal}{Nucl. Phys. B} \textbf{\bibinfo{volume}{909}},
  \bibinfo{pages}{243 } (\bibinfo{year}{2016}).

\bibitem[{\citenamefont{Chen and L\"u}(2018)}]{mypaper6}
\bibinfo{author}{\bibfnamefont{X.}~\bibnamefont{Chen}} \bibnamefont{and}
  \bibinfo{author}{\bibfnamefont{X.}~\bibnamefont{L\"u}},
  \bibinfo{journal}{Phys. Rev. D} \textbf{\bibinfo{volume}{97}},
  \bibinfo{pages}{114005} (\bibinfo{year}{2018}).

\bibitem[{\citenamefont{Uehara et~al.}(2010)}]{X39151}
\bibinfo{author}{\bibfnamefont{S.}~\bibnamefont{Uehara}} \bibnamefont{et~al.}
  (\bibinfo{collaboration}{Belle Collaboration}), \bibinfo{journal}{Phys. Rev.
  Lett.} \textbf{\bibinfo{volume}{104}}, \bibinfo{pages}{092001}
  (\bibinfo{year}{2010}).

\bibitem[{\citenamefont{Lees et~al.}(2012)}]{Y39404}
\bibinfo{author}{\bibfnamefont{J.~P.} \bibnamefont{Lees}} \bibnamefont{et~al.}
  (\bibinfo{collaboration}{BABAR Collaboration}), \bibinfo{journal}{Phys. Rev.
  D} \textbf{\bibinfo{volume}{86}}, \bibinfo{pages}{072002}
  (\bibinfo{year}{2012}).

\bibitem[{\citenamefont{Aaltonen et~al.}(2009)}]{Y39403}
\bibinfo{author}{\bibfnamefont{T.}~\bibnamefont{Aaltonen}} \bibnamefont{et~al.}
  (\bibinfo{collaboration}{CDF Collaboration}), \bibinfo{journal}{Phys. Rev.
  Lett.} \textbf{\bibinfo{volume}{102}}, \bibinfo{pages}{242002}
  (\bibinfo{year}{2009}).

\bibitem[{\citenamefont{Esposito et~al.}(2017)\citenamefont{Esposito, Pilloni,
  and Polosa}}]{mutlistate}
\bibinfo{author}{\bibfnamefont{A.}~\bibnamefont{Esposito}},
  \bibinfo{author}{\bibfnamefont{A.}~\bibnamefont{Pilloni}}, \bibnamefont{and}
  \bibinfo{author}{\bibfnamefont{A.~D.} \bibnamefont{Polosa}},
  \bibinfo{journal}{Phys. Rep.} \textbf{\bibinfo{volume}{668}},
  \bibinfo{pages}{1} (\bibinfo{year}{2017}).

\bibitem[{\citenamefont{Burden et~al.}(1997)\citenamefont{Burden, Qian,
  Roberts, Tandy, and Thomson}}]{BSE:Roberts1}
\bibinfo{author}{\bibfnamefont{C.~J.} \bibnamefont{Burden}},
  \bibinfo{author}{\bibfnamefont{L.}~\bibnamefont{Qian}},
  \bibinfo{author}{\bibfnamefont{C.~D.} \bibnamefont{Roberts}},
  \bibinfo{author}{\bibfnamefont{P.~C.} \bibnamefont{Tandy}}, \bibnamefont{and}
  \bibinfo{author}{\bibfnamefont{M.~J.} \bibnamefont{Thomson}},
  \bibinfo{journal}{Phys. Rev. C} \textbf{\bibinfo{volume}{55}},
  \bibinfo{pages}{2649} (\bibinfo{year}{1997}).

\bibitem[{\citenamefont{Maris et~al.}(1998)\citenamefont{Maris, Roberts, and
  Tandy}}]{BSE:Roberts3}
\bibinfo{author}{\bibfnamefont{P.}~\bibnamefont{Maris}},
  \bibinfo{author}{\bibfnamefont{C.~D.} \bibnamefont{Roberts}},
  \bibnamefont{and} \bibinfo{author}{\bibfnamefont{P.~C.} \bibnamefont{Tandy}},
  \bibinfo{journal}{Phys. Lett. B} \textbf{\bibinfo{volume}{420}},
  \bibinfo{pages}{267} (\bibinfo{year}{1998}).

\bibitem[{\citenamefont{Ivanov et~al.}(1999)\citenamefont{Ivanov, Kalinovsky,
  and Roberts}}]{BSE:Roberts4}
\bibinfo{author}{\bibfnamefont{M.~A.} \bibnamefont{Ivanov}},
  \bibinfo{author}{\bibfnamefont{Y.~L.} \bibnamefont{Kalinovsky}},
  \bibnamefont{and} \bibinfo{author}{\bibfnamefont{C.~D.}
  \bibnamefont{Roberts}}, \bibinfo{journal}{Phys. Rev. D}
  \textbf{\bibinfo{volume}{60}}, \bibinfo{pages}{034018}
  (\bibinfo{year}{1999}).

\bibitem[{\citenamefont{Ivanov et~al.}(2007)\citenamefont{Ivanov, K\"orner,
  Kovalenko, and Roberts}}]{BSE:Roberts5}
\bibinfo{author}{\bibfnamefont{M.~A.} \bibnamefont{Ivanov}},
  \bibinfo{author}{\bibfnamefont{J.~G.} \bibnamefont{K\"orner}},
  \bibinfo{author}{\bibfnamefont{S.~G.} \bibnamefont{Kovalenko}},
  \bibnamefont{and} \bibinfo{author}{\bibfnamefont{C.~D.}
  \bibnamefont{Roberts}}, \bibinfo{journal}{Phys. Rev. D}
  \textbf{\bibinfo{volume}{76}}, \bibinfo{pages}{034018}
  (\bibinfo{year}{2007}).

\bibitem[{\citenamefont{Chen et~al.}(2013)\citenamefont{Chen, Liu, Shi, and
  L\"u}}]{mypaper3}
\bibinfo{author}{\bibfnamefont{X.}~\bibnamefont{Chen}},
  \bibinfo{author}{\bibfnamefont{R.}~\bibnamefont{Liu}},
  \bibinfo{author}{\bibfnamefont{R.}~\bibnamefont{Shi}}, \bibnamefont{and}
  \bibinfo{author}{\bibfnamefont{X.}~\bibnamefont{L\"u}},
  \bibinfo{journal}{Phys. Rev. D} \textbf{\bibinfo{volume}{87}},
  \bibinfo{pages}{065013} (\bibinfo{year}{2013}).

\bibitem[{\citenamefont{Luri\'{e}}(1968)}]{Mandelstam}
\bibinfo{author}{\bibfnamefont{D.}~\bibnamefont{Luri\'{e}}},
  \emph{\bibinfo{title}{{Particles and Fields}}}
  (\bibinfo{publisher}{Interscience Publishers, New York},
  \bibinfo{year}{1968}).

\bibitem[{\citenamefont{Neubert}(1994)}]{hqet}
\bibinfo{author}{\bibfnamefont{M.}~\bibnamefont{Neubert}},
  \bibinfo{journal}{Phys. Rep.} \textbf{\bibinfo{volume}{245}},
  \bibinfo{pages}{259} (\bibinfo{year}{1994}).

\bibitem[{\citenamefont{Patrignani et~al.}(2016)}]{PDG2016}
\bibinfo{author}{\bibfnamefont{C.}~\bibnamefont{Patrignani}}
  \bibnamefont{et~al.} (\bibinfo{collaboration}{Particle Data Group}),
  \bibinfo{journal}{Chin. Phys. C} \textbf{\bibinfo{volume}{40}},
  \bibinfo{pages}{100001} (\bibinfo{year}{2016}).

\end{thebibliography}

\end{document}